\documentclass[twocolumn,showpacs,preprintnumbers,amsmath,amssymb]{revtex4}

\usepackage{graphicx}
\usepackage{dcolumn}
\usepackage{bm}

\begin{document}
\preprint{APL/By H. Y. Wang}

\title{Modeling of Perpendicularly Driven Dual-Frequency Capacitively Coupled Plasma\\}

\author{Hong-Yu Wang$^{1,2}$(ÍõºçÓî)}
\author{Wei Jiang$^1$(½ªÎ¡)}
\author{Zhen-Hua Bi$^1$(±ÏÕñ»ª)}
\author{You-Nian Wang$^1$(ÍõÓÑÄê)}
\email{ynwang@dlut.edu.cn} \affiliation{$^1$School of Physics and
Optoelectronic Technology, Dalian University of Technology, Dalian,
116024, P.R.China\\$^2$Department of Physics, Anshan Normal
University, Anshan,
 114007, P.R.China}

\date{\today}

\begin{abstract}
We proposed an altered configuration for dual-frequency (DF)
capacitively coupled plasmas (CCP). In this configuration, two pairs
of electrodes are arranged oppositely, and the discharging is
perpendicularly driven by two rf sources. With
Particle-in-cell/Monte Carlo method, we have demonstrated this
configuration can remove the harmful electromagnetic and DF coupling
effects in conventional DF-CCP.
\end{abstract}

\maketitle

Dual-frequency (DF) capacitively coupled plasmas (CCP) are commonly
used as etching and deposition devices in the microelectronics, flat
panel display and solar cells industries
\cite{Lieberman05,Makabe06}. Compare to the other two sources,
inductive coupled plasma (ICP) and electron cyclotron resonance
(ECR) discharges, CCP can produce uniform plasma over larger areas.
In typical DF-CCP \cite{Wakayama03,Boyle04}, the two rf source with
different frequencies are applied to the same electrode or the
opposite two electrodes. The high frequencies (hf) source controls
the plasma density while the low frequency (lf) source controls the
ion flux and ion energy and angular distribution (IEDs and IADs).
Quasi-independent control of plasma density and IEDFs is the main
merit of DF-CCP over single-frequency CCP. Therefore DF-CCP has been
widely used in state-of-art industry reactors. To achieve more
flexibilities, some variants of DF-CCP, such as adopting very high
frequency (VHF)\cite{Rauf09}, applying additional dc source
\cite{Kawamura08} and series resonance CCP \cite{Mussenbrock08},
have received intense investigation recently.

High etching rates, uniformity, anisotropy, selectivity and low
dielectric damage are the essential requirements of the CCP etching
devices. There is no perfect solutions to satisfy all the
requirements above simultaneously, one must make tradeoffs. Although
successful in practice, DF-CCP still expose some problems. The first
one is the electromagnetic (EM) effect\cite{Abdel03,Gans06,Ahn08},
which will occur when very high frequency (VHF) sources (typically
$>60$MHz) are used, and therefore the excitation wavelength
$\lambda$ is comparable to the reactor radius. Due to the standing
wave effect \cite{Chabert05}, the merit of plasma uniformities over
large areas will be broken. The second one is hf-lf source coupling
effects \cite{Lee05}. When only increasing the hf voltage, the
density increases and the sheath thickness decreases, thus more ions
will appear in the high energy end of IEDs. This is not desired
because dielectric damage may be introduced. But when only
increasing the lf voltage, the sheath thickness increases and the
bulk plasma lengths decrease and thus the density decreases. This is
also not desired because of the reduced etching rate. Some
configurations, like shaped electrodes \cite{Sansonnens03} and
asymmetrical discharges \cite{Donko09}, are proposes to mitigate
above harmful effects.
\begin{figure}
\includegraphics[scale=0.7]{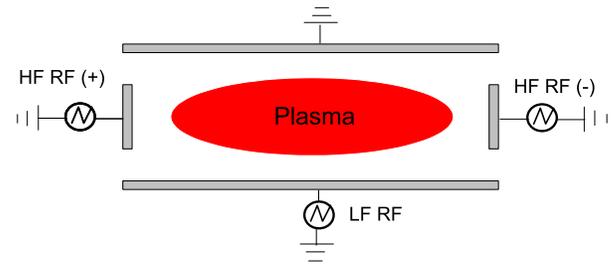}
\caption{\label{Configure} Schematic of the perpendicularly driven
DF-CCP.}
\end{figure}

In this letter, we propose an altered configuration for DF-CCP. Not
like the conventions disk-like cylindrical configuration
\cite{Lieberman05}, here the discharge chamber is a
three-dimensional (3D) flat regular hexahedron, whose
cross-sectional schematic is shown in Fig.\ref{Configure}. Two pairs
of rectangular electrodes are arranged oppositely, in the left/right
surface and the top/bottom surface, respectively. In the front/back
surface (not shown in the figure) and in the gaps between the
electrode, thick dielectric insulators are placed to confine the
plasmas. The discharging is perpendicularly driven by two rf
sources. The smaller pairs (left and right) electrodes (LE and RE)
are powered by the same hf sources, but with the phase difference of
$ \pi$. This can be realized with a opposite phase power divider,
for example, a balun \cite{Vizmuller95} applied between the rf
source and the devices. The bottom electrode (BE) is powered by the
lf source and the top electrode (TE) is grounded. The wafer is
placed on the BE. We will demonstrate that this configuration can
remove above problems.

In order to study such a configuration, we have adopted direct
implicit Particle-in-cell/Monte Carlo (PIC/MC) algorithms
\cite{Langdon82} in 2D planar geometry. Although this configuration
is inherently 3D, due to the confinement by the insulators, the
plasma is uniform between the side-wall insulators and thus 2D model
is sufficient, if the front/back spacing is large. The details of
the algorithms are described elsewhere \cite{Wang09}, therefore we
will only present the simulation parameters here. The BE and TE
lengths are $X=8$cm, and the LE and RE lengths are $Y=1.5$cm,
therefore we have the four gaps of $0.25$cm. The hf sources of
$\omega_{hf}=60$MHz and $V_{hf}=50$V or $100$V are applied to LE and
RE, with the phase difference of $ \pi$. The lf sources of
$\omega_{lf}=2$MHz and $V_{lf}=50$V or $100$V are applied to BE. In
the gaps, the instant potentials are linearly interpolated. Argon
gas is used with the pressure of $10$mTorr and temperature of
$300K$. We consider elastic, excitation and ionization collisions
for electrons and elastic and charge transfer collisions for Ar$^+$
ions, respectively. Square cells are used, thus X direction is
uniformly divided to 256 cells and Y direction divided to 64 cells.
The space and time steps are fixed to all simulations, $\Delta
x=0.02/64$m, $\Delta t_e=\Delta t_i=0.5\times10^{-10}$s.  All
results are given by averaging over one lf period after reaching
equilibrium of 1000 rf periods. Because the diodes areas are equal
for X / Y separately and only the ideal voltage sources in the
external circuit, we did not consider the external circuit and the
self-biasing for simpleness.

\begin{figure}
\includegraphics[scale=0.7]{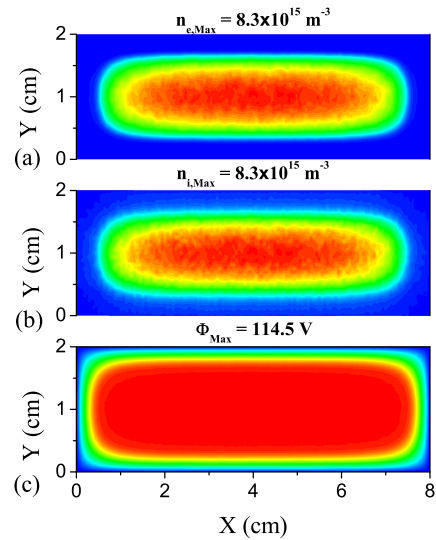}
\caption{\label{Phi} Average (a)electron density $n_e$, (b)ion
density $n_i$ and (c) potential $\Phi$ for the case of $V_{hf}=100$V
and $V_{lf}=50$V.}
\end{figure}

The average potential, electron and ion density for the case of
$V_{hf}=100$V and $V_{lf}=50$V are shown in Fig.\ref{Phi}. Near each
electrode, there is a sheath, therefore the densities are in an
elliptical profile and the maximum value is in the center. The
sheathes are symmetric in both directions, but the average sheath thicknesses are
larger for TE and BE than that for LE and RE. This is a natural
result, since when the density is fixed, the sheath thickness
inversely scales with the frequency\cite{Lieberman05}.

\begin{figure}
\includegraphics[scale=0.6]{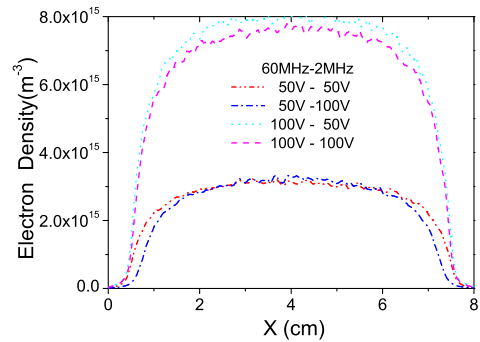}
\caption{\label{Ne_X} Cross-sectional profiles of electron density
$n_e$ at $Y=1$cm for different voltages.}
\end{figure}

We plotted the cross-sectional profiles of $n_e$ for different
voltages in Fig.\ref{Ne_X}. The plasma density is mainly determined
by the hf source. When increasing the $V_{hf}$ by two times, the
electron densities also increase by a factor of larger than 2. While
when increasing $V_{lf}$, the electron densities are nearly
unchanged. The cross-sectional profiles are flat in the center. As
we have mentioned, in conventional DF-CCP, when increasing the lf
voltage, the bulk plasma length will decrease and thus the density
will decrease\cite{Lee05}. In this present configuration, there is no
such a effect, since the sheath thicknesses are decoupled.

\begin{figure}
\includegraphics[scale=0.6]{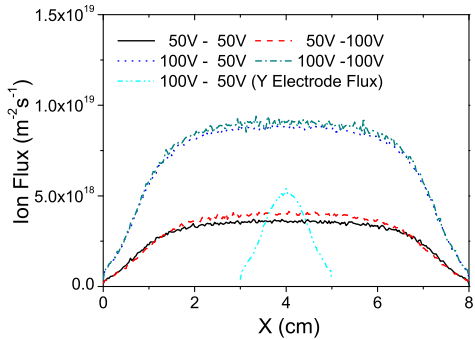}
\caption{\label{Flux} Ion flux onto the BE for different voltages.
We also plotted the flux to LE for comparison}
\end{figure}

The ion flux to BE is presented in Fig.\ref{Flux}. Similar to the
density, the flux is mainly determined by the hf source. When
increasing the $V_{hf}$ by two times, the flux also increase by a
factor of larger than 2. While $V_{lf}$ has no significant effects
on the flux. Compare the convention CCP driven by $13.56$MHz source,
the flux is much larger even the density value is close
\cite{Wang09}. In industry etching and deposition devices, it is
critical to make the ion flux to the electrode be uniform over
larger areas. As can be seen, over large distance (about 6cm), the
ion flux is uniform. In VHF CCP, due to the finite wave lengths
effect as we have mentioned, the plasma density and ion flux will
not be uniform over the wafer. In this configuration, the wafer can
be placed on X plate while the hf sources are in Y direction,
therefore the harmful EM effects are naturally removed.

\begin{figure}
\includegraphics[scale=0.6]{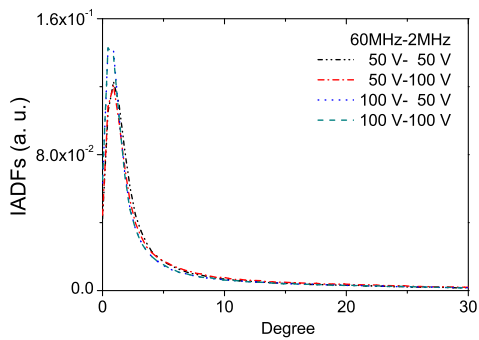}
\caption{\label{IADF} Ion angle distributions (IADs) for different
voltages.}
\end{figure}

In industry devices, it is also highly desired that the ions are
anisotropic. The ion angle distributions (IADs) for different
voltages are depicted in Fig.\ref{IADF}. Whatever the rf voltage is,
most ions has the angle of several degrees. This means the ions
still keep anisotropic in this configuration.

The last issue in this configuration is IEDs. The IEDs is influenced
not only by rf frequency and voltage, but also by the ion transit
time $\tau_i$. In conventional DF-CCP, if one increases $V_{hf}$
while keeping the $V_{lf}$ constant, the plasma density will
increases and the sheath thickness will decrease. Therefore $\tau_i$
will decreases and the profiles of the IEDFs will correspondingly
shift to the higher energy tails in some cases. This will result
dielectric damage as we have mentioned. Furthermore, the etching
uniformity also requires the IEDs being unchanged at different
position on the electrodes.

\begin{figure}
\includegraphics[scale=0.6]{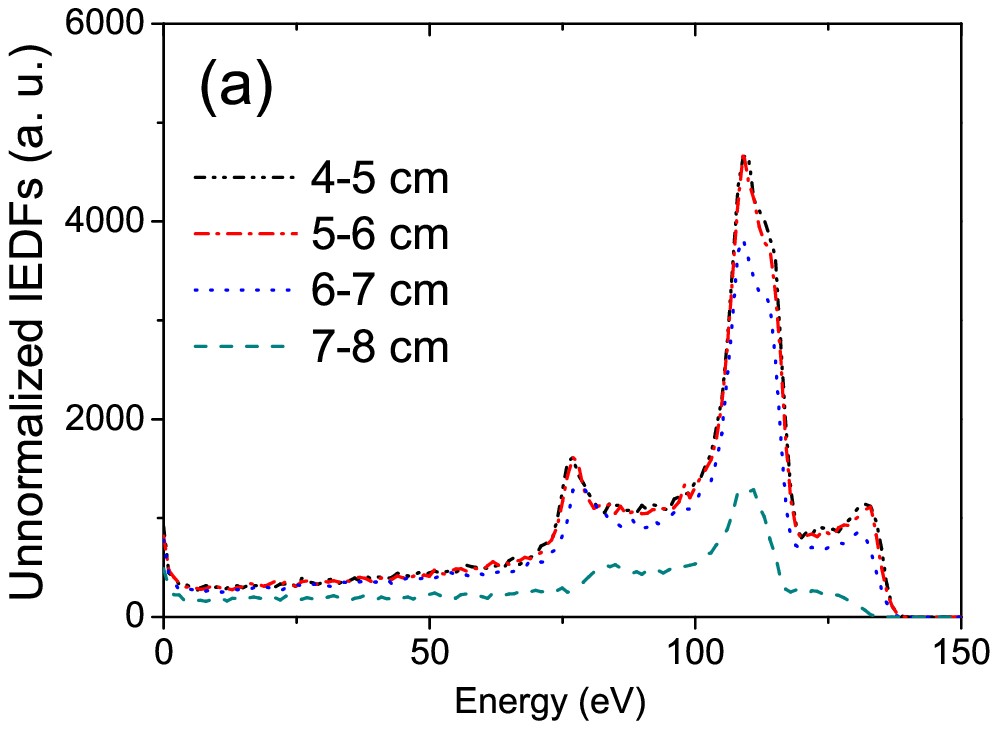}
\includegraphics[scale=0.6]{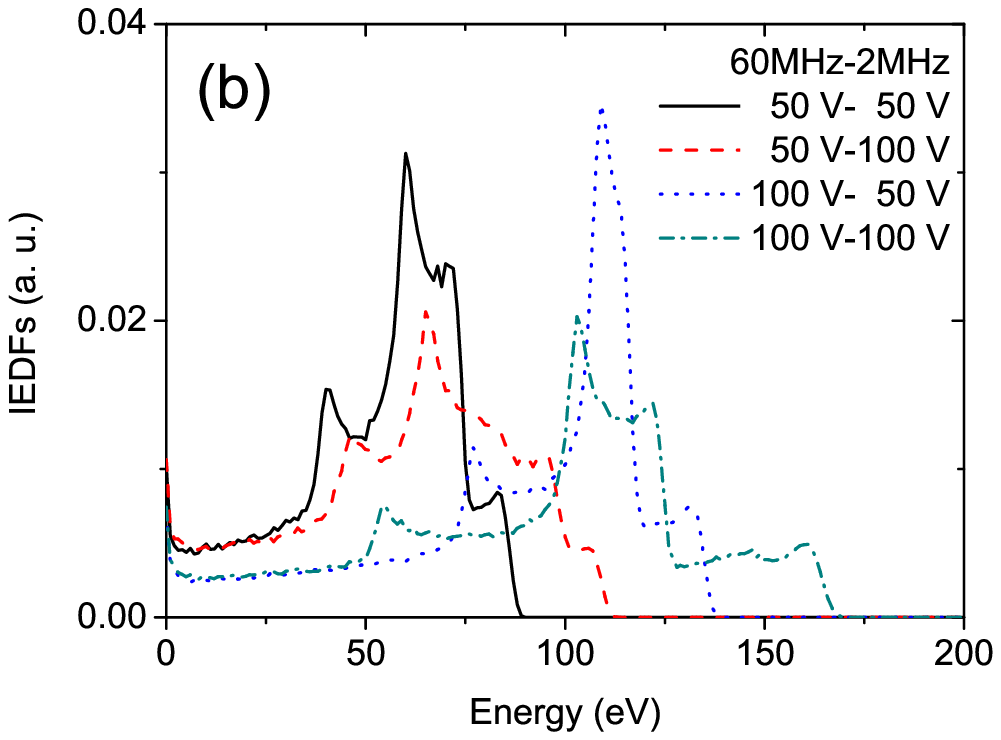}
\caption{\label{IEDF} (a)Ion energy distributions (IEDs) for
different positions for the case of $V_{hf}=100$V and $V_{lf}=50$V;
(b)Ion energy distributions (IEDs) for different rf voltages.}
\end{figure}

We plotted unnormalized IEDs for different positions for the case of
$V_{hf}=100$V and $V_{lf}=50$V in Fig.\ref{IEDF}(a). Here the IEDs
are sampled over four 1cm segments beginning at the center of the
BE. It can be seen that the shape of IEDs is nearly unchanged over
3cm. Note here the slight amplitude difference is from the flux
difference (Fig.\ref{Flux}). Only in the electrode edge, the shape
is different.  The reason is the ions always response to the average
electric field. In the sheath near the X electrode, the ions are
mainly accelerated by the field $E_y$ produced by the lf source.

We showed the IEDs over entire BE for different voltages in
Fig.\ref{IEDF}(b), since the IEDs are similar at different position.
There are clearly four peaks in all the IEDs, and the peaks can be
divided in to two pairs. Each pair of the peaks seems to be produced
by the hf or lf source solely, namely, the lower peaks pair are
produced by the lf source and the high peak are produced by the hf
source. For we still have $\omega_{rf}>\omega_i$, the center energy
of IEDs $V_i$ still obey the simple estimation of
$0.4(V_{lf}+2V_{hf})$, where 2 denote the two hf sources. Unlike the
conventional DF-CCP, the IEDs do not shift to the higher energy
tail. If $V_{lf}>>V_{hf}$, the IEDs will be solely determined by the
lf source.

In summary, we have proposed an altered configuration of DF-CCP,
which is in 3D flat regular hexahedron shape and is perpendicularly
driven by two rf sources. The plasma density and ion flux are solely
determined by the hf source, and are uniform over larger area,
without the harmful EM and DF coupling effect. At the same time, the
IEDs are mainly determined by the lf source, there are no excessive
high energy ions in the tail of the IEDs, which will avoid the
dielectric damage. In practical devices, the geometry is 3D, but the
qualitative results here will not change and one may even adopted a
triple frequency configuration. If X length is very large, rf
breakdown laws by Lisovskiy \cite{Lisovskiy08} should be considered
for the reactors design.

This work was supported by the National Natural Science Foundation
of China (No.10635010).


\begin{thebibliography}{99}                                                                                               %
\bibitem{Lieberman05} Lieberman M A and Lichtenberg A J Principles of Plasma
 Discharges and Materials Processing 2nd edn 2005. (New York: Wiley)
\bibitem{Makabe06} T. Makabe and  Z. L. Petrovic Plasma Electronics: Applications in
Microelectronic Device Fabrication 2006. (New York: Taylor and
Francis Group)
\bibitem{Wakayama03} G. Wakayama and K. Nanbu, IEEE Trans. Plasma Sci. \textbf{31} 638 (2003).
\bibitem{Boyle04} P. C. Boyle,  A. R. Ellingboe, and M. M. Turner Plasma Sources Sci. Technol. \textbf{13} 493 (2004).
\bibitem{Rauf09} Rauf S, Kenney J and Collins K, J. Appl. Phys. \textbf{105} 103301 (2009).
\bibitem{Kawamura08} E. Kawamura, M. A. Lieberman, and A. J. Lichtenberg, Plasma Sources Sci. Technol. \textbf{17} 045002 (2008).
\bibitem{Mussenbrock08} T. Mussenbrock, R. P. Brinkmann, M. A. Lieberman, A. J. Lichtenberg, and E. Kawamura, Phys. Rev. Lett. {\bf 101}, 085004 (2008).
\bibitem{Abdel03} E. Abdel-Fattah and H. Sugai, Appl. Phys. Lett. \textbf{83}, 1533 (2003).
\bibitem{Gans06} T. Gans, J. Schulze, D. O'Connell, U. Czarnetzki, R. Faulkner, A. R. Ellingboe, and M. M. Turner, Appl. Phys. Lett. \textbf{89}, 261502 (2006).
\bibitem{Ahn08} S. K. Ahn and H. Y. Chang, Appl. Phys. Lett. \textbf{93}, 031506 (2008).
\bibitem{Chabert05} Chabert P, Raimbault J L, Levif P, Rax J M and Lieberman M A 2005 Phys. Rev. Lett. \textbf{95}, 205001
\bibitem{Lee05} J. K. Lee, O. V. Manuilenko, N. Y. Babaeva., H. C. Kim, and J. W. Shon, Plasma Sources Sci. Technol. \textbf{14} 89 (2005).
\bibitem{Sansonnens03} L. Sansonnens and J. P. M. Schmitt, Appl. Phys. Lett. \textbf{82}, 182 (2003).
\bibitem{Donko09} Z. Donko, J. Schulze, B. G. Heil, and U. Czarnetzki, J. Phys. D: Appl. Phys. \textbf{42} 025205 (2009).
\bibitem{Vizmuller95} P. Vizmuller, RF Design Guide - System, Circuits and
Equations, 1996. (Norwood: Artech House)
\bibitem{Langdon82} A. B. Langdon, B. I. Cohen and A. Friedman J. Comput. Phys. \textbf{46} 15 (1982).
\bibitem{Wang09} H. Y. Wang, W. Jiang and Y. N. Wang, Implicit and electrostatic Particle-in-cell/Monte Carlo model in
two-dimensional and axisymmetric geometry, submitted to Plasma
Sources Sci.Technol.
\bibitem{Lisovskiy08} V. Lisovskiy, J. P. Booth, K. Landry, D. Douai, V. Cassagne, and V. Yegorenkov, EPL, \textbf{82} 15001 (2008).
\end{thebibliography}
\end{document}